\documentclass[prb,preprint,superscriptaddress]{revtex4}
\usepackage[latin1]{inputenc}
\usepackage{amsmath}
\usepackage{verbatim}
\usepackage{amssymb}
\usepackage{bbm}
\usepackage{graphicx}

\begin{document}
\title{Two universal physical principles shape the power-law statistics of real-world networks }
\author{Tom Lorimer, Florian Gomez, Ruedi Stoop}
\email{ruedi@ini.phys.ethz.ch}
\affiliation{Institute of Computational Science and
Institute of Neuroinformatics, University of Zurich and ETH Zurich, Winterthurerstrasse 190, 8057 Zurich, Switzerland}

\date{\today}

\begin{abstract}

The study of complex networks has pursued an understanding of macroscopic behavior by focusing on power-laws in microscopic observables. Here, we uncover two universal fundamental physical principles that are at the basis of complex networks generation. These principles together predict the generic emergence of deviations from ideal power laws, which were previously discussed away by reference to the thermodynamic limit. Our approach proposes a paradigm shift in the physics of complex networks, toward the use of power-law deviations to infer meso-scale structure from macroscopic observations. 

\end{abstract}

\maketitle

\section*{Introduction}
A recent seminal discovery elucidated that in nature a simple physical principle rules  often the growth of `random networks'. The so called preferential attachment (`the rich get richer') rule leads to complex networks that have properties contrasting those predicted from classical random network theory \cite{Boccaletti_review,barabasi_review,shlomo_book,barabasi_albert}. 
A fundamental universality principle of physics must be held responsible for this change of paradigm. The preferential attachment principle expresses in our interpretation that for the formation of ensembles, attractive forces that are generally valid over decades of spatial extensions are required (that in physics may involve mass, charge, e.g.). It is this principle that generates the celebrated power laws observed in the distribution of mesoscopic network indicators, such as network degree, connectivity weight  \cite{stanley1,stanley2,dorogovtsev_mendes_wordwebs, Boccaletti2}, or neuronal avalanche size \cite{hans1, herman1, herman2}. 
A {\it second} fundamental universality principle of physics is, however, active at the same time, that has passed unnoticed so far. It is the fact that real-world connectivity requires space, and that this space is limited. The question that we address in our work is what the traces of this principle will be, during network formation and regarding the final network. This question has not been answered so far.

\section*{Generic network building algorithm}

\begin{figure}[h!!!!!!!!!!!!!!!!!!!!!!]
  \centering
  	\includegraphics[width=80mm]{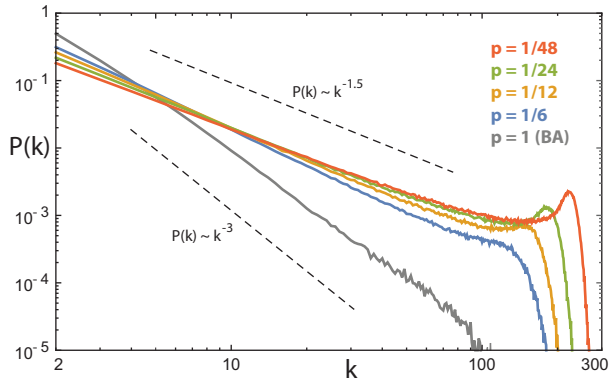}
    \caption{Characteristic degree distributions from the two key principles (for different values of parameter $p$ and fixed parameter $m=2$; the effect of $m$ is exhibited in Fig. \ref{fig:varyp2} and Fig. \ref{fig:varym}). Network size $t=10^3$ nodes, mean of $10^3$ realizations. Dashed lines: power-law visual guides. The effect is most saliently expressed for exponents $<2$, occurring often in gene or protein networks.}
    \label{fig:varyp}
    \end{figure}

To study this question, we consider a novel generic network building algorithm (our 'primary model') that implements both principles at the most basic level as follows. We start from a connected network of $N_0$ nodes. With probability $p$, an `outside' node, from a finite set of available nodes, is added; alternatively, with probability $1-p$, an attempt is made to construct an 'inside' edge (see below). If an outside node is added, the new node joins the network by $m$ edges, where the target nodes are sampled according to their degree $k$ (i.e. $\propto k$), following preferential attachment. For an inside edge, two nodes are independently chosen along preferential attachment (i.e., proportional to the degree they have). If the two chosen nodes are not identical and not already connected, an edge is established. In this way, the algorithm's second alternative expresses the second fundamental principle in terms of  an 'edge saturation' (at a level defined by $p$ and $m$, implemented right from the start of the network's growth). The process stops if the set of available nodes is depleted. The algorithm generates undirected topological networks of arbitrary size, void of loops and multiple-edges; examples will be discussed later. Fig. \ref{fig:varyp} shows the stereotypical degree distribution obtained in this way, exhibiting an extended power-law part of the distribution terminated by a hump (that, upon the network's growth, moves towards larger degrees, until the process is stopped by node depletion, cf. Fig. \ref{fig:saturation}b).

\section*{Network properties}

\begin{figure}[h!!!!!!!!!!!!!!!!!!!!!!]
  \centering
  	\includegraphics[width=88mm]{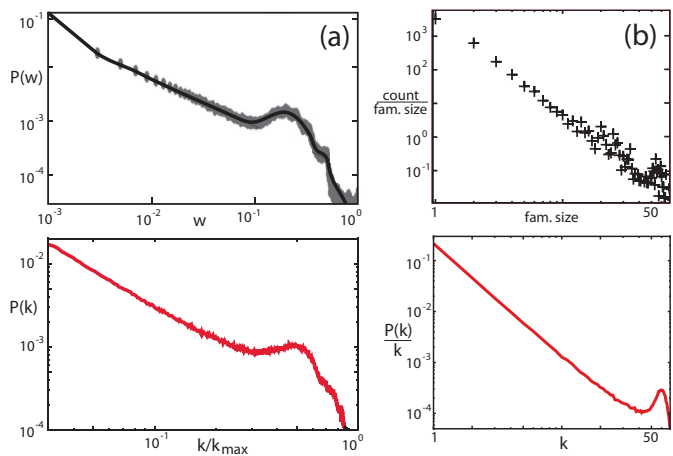}
    \caption{Typical weight and degree distributions, respectively, from experiments, and their qualitative modeling (black: experimental, red: simulation data). a) Network of synchronizing linear phase oscillators (network weight distribution during synchronization) \cite{Boccaletti2}. b) Gene family for {\it S. cerevisiae} \cite{Camacho} (family size distribution). For the modeling, different 
    $(p,m)$-models were superimposed for a).}
    \label{fig:varyp0}
    \end{figure}

\begin{figure}[h!!!!!!!!!!!!!!!!!!!!!!]
  \centering
  	\includegraphics[width=88mm]{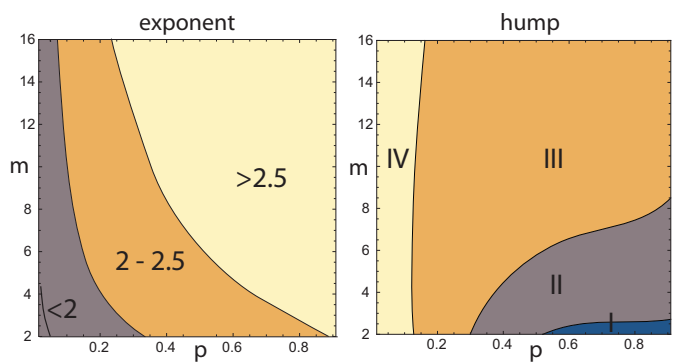}
    \caption{Modeling guidelines: Phase diagram of the humped power law's exponent and 'humpiness' on
     local parameters $(p,m)$ (see text). 
Domains of humpiness: I) not resolvable, II minor, III significant, IV salient. Guided by the power-law paradigm, investigations have mostly focused on examples from domains I and II. Network sizes: $t=10^3$.}
    \label{fig:varyp2}
    \end{figure}

While we observe a wide-spread activity to find power-law distributions in all areas of physics, we emphasize that based on the fundamental ingredients necessary in the network building process, only in rare cases neat power laws will be found. 
Examples of experimental data with the deviations that our key principles predict are shown in Fig. \ref{fig:varyp0}. While our real-world examples are often related to biology (mostly because of the great availability of the underlying data, and because of the greater simplicity of the examples), all of our arguments are immediately transferable to physical situations where previous analysis has generally stopped at the preferential attachment level. Our analysis now provides guidelines for inferring from macroscopic measurements the microscopic properties that dominate network growth (cf. Fig. \ref{fig:varyp2}, where the 'humpiness' of the distribution $P(k)$ was evaluated as the deviation from the power law $p(k)$ excluding the hump, as $(P(k) - p(k))/p(k)$). This provides an important input for the modeling of real world systems (see, e.g., the Drosophila network example discussed below). By superposition of prototypes with different $p$ and $m$ parameters, more general hump structures can be generated (Fig. \ref{fig:varyp0}). This mechanism provides an as yet unexplored link between the macro- and meso-scales that can be invaluable for both the modeling and the further analysis of real-world systems.

In contrast to preferential attachment networks (cf. \cite{GrossPRL}), a network generated along the two fundamental physical principles embodied in our primary model, will not be necessarily sparse (this would imply a power-law exponent $>2$ , cf. Fig. \ref{fig:varyp}). Moreover, also Dorogovtsev and Mendes' modified preferential attachment algorithm with its double regimes of power-law behavior \cite{dorogovtsev_mendes_wordwebs} deviates from the fundamental principles that we have worked out. That model uses a second internal linking process that is always successful in making new connections. In our case it is exactly the edge connection failures (by edge saturation) that define the network structure. Whereas the rate of internal linking in their algorithm accelerates with the network size, our approach does not share this property. Moreover, the network structures that we obtain depend primarily on parameter $p$ and the obtained distributions are generally unaffected by the network's initial condition (in contrast to Refs. \cite{dorogovtsevPRE,waclaw,guimaraes}).

\begin{figure}[!b!!!!!!!!!!!!!!!!!!!!!!]
\centering
  	\includegraphics[width=87.5mm]{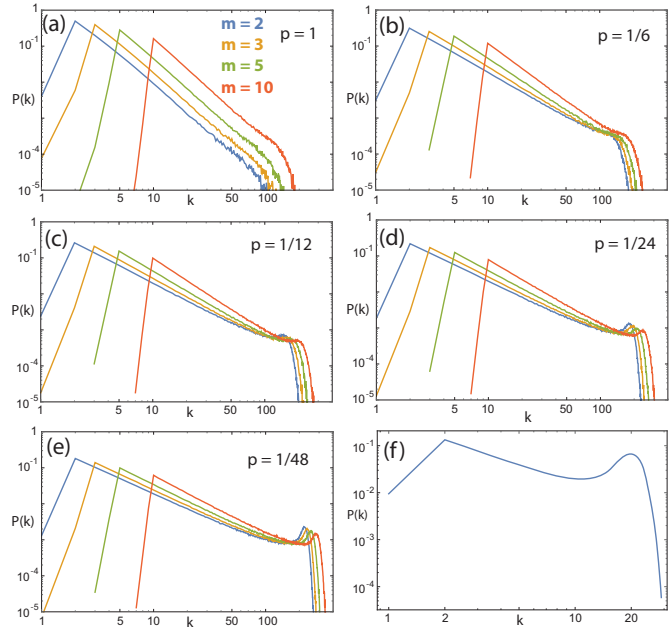}
    \caption{\label{fig:varym} a)-e) Choice of  $m$ on network degree distribution, for different values of $p$ (network size $t=10^3$ nodes, mean of $10^3$ realizations). Increasing $m$ for $p << 1$ increases the influence of the first term in Eq. (3), which increases the exponent by pushing the primary model towards the preferential attachment model. f) Real-world example: Drosophila courtship network's degree distribution (corresponding to the full line in Fig. \ref{fig:varym2}). Degrees $k < m$ have small probability.}
  \end{figure}

\begin{figure}[t!!!!!!!!!!!!!!!!!!!!!!]
\centering
  	\includegraphics[width=85mm]{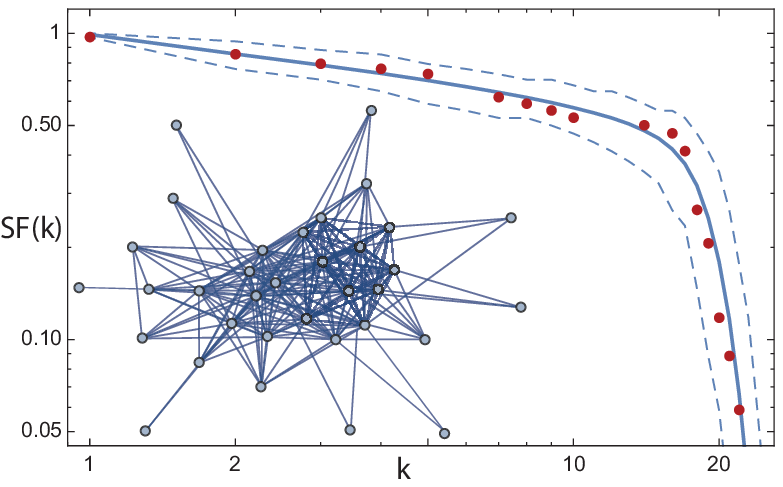}
    \caption{\label{fig:varym2} Drosophila courtship language network degree distribution. a) Survival function $SF(k):=1-CDF(k)$, where $CDF$ is the cumulative distribution function (red dots: original data). Solid line: means, dashed lines: $0.05$ quantiles, from 1000 realizations of our network growth algorithm ($N=34$, $p=\frac{1}{27}$, $m=2$). Inset: mapped-out \emph{Drosophila} language network.}
    \end{figure}

The modeling of biological networks containing a small number of nodes only, is a particular challenge. The example of Drosophilas's courtship network, a network that is  built on observable irreducible acts of body language \cite{stoop4,stoop5} (cf. Figs. \ref{fig:varym} and \ref{fig:varym2}) illustrates that our approach also successfully masters this challenge (a further discussion of this example is given towards the end of the paper). 

\section*{Statistical modeling}

To better understand how the statistical properties and in particular, saturation, emerge from the model, we focus on a semi-analytical growth description, in which the natural time step $t$ is the addition of one node to the network.
The degree distribution from a network growth algorithm is usually determined from a differential equation that describes the rate of addition of new edges to a given node, as a function of the time $s$ at which the node has joined the network \cite{dorogovtsev_mendes_book}, i.e. $\frac{\partial k(s,t)}{\partial t} = f(k,s,t).$
For our algorithm, the topological constraint on the addition of inside edges implies that $\frac{\partial k(s,t)}{\partial t}$ can not be determined analytically from the single node information $f(k,s,t)$, but requires the full pairwise connection information of the network encoded in the adjacency matrix at time $t$, $A_t$, i.e. $$\frac{\partial k(s,t)}{\partial t} = f(k,s,t,A_t).$$
To work around this complication, we make the following ansatz. We suppose that the probability of failure while trying to add an inside edge $(i,j)$ to an already chosen node $i$, can be expressed by a mean field `saturation' function $F(k,t)$ in terms of the degree $k$ of node $i$. Furthermore, suppose that the total number of edges present in the network at time $t$ can be approximated by $K(t)$.
$F(k,t)$ is then defined as the average probability of a node with degree $k$, to be already connected to a second node $j$ chosen with $P \propto k_j$.  Thus,
\begin{equation}
F(k,t):=\bigl<F_i(t)\bigr>_{k_i=k}\,\,\,\,\,\,,
\end{equation}
where $F_i(t)$ is the probability that node $i$ with degree $k_i$, is already connected to node $j$.  
$F_i(t)$ has then the form
\begin{equation}
F_i(t):=\frac{k_i(t) + \sum_{(i,j)\in E(t)}{k_j(t)}}{\sum_j{k_j(t)}},
\end{equation}  
where $k_i(t)$ accounts for the case where node $i$ would be chosen twice, and the second term is the degree-weighted sum over the nodes to which node $i$ is already connected ($E(t)$ denotes the network's set of edges). 

Using this approximation, we can express our algorithm by the rate of addition of new edges to a node of degree $k(s,t)$ as
\begin{equation}
\frac{\partial k(s,t)}{\partial t} = \frac{m k(s,t)}{2 K(t)}\; + \;\frac{1-p}{p}\,\frac{k (s,t)}{K(t)}[1-F(k,t)]\,.
\label{eqn:rate}
\end{equation}
In this case, the network grows out from a connected network of  $N_0$ nodes, with $k(s,s) \approx m$ as the initial condition.  The first term on the right hand side of Eq. (\ref{eqn:rate}) describes the increase in $k$ due to connection to outside nodes, and the second term describes the addition of inside edges.  The whole equation has been rescaled by $\frac{1}{p}$ (canceling the $p$ in the first term's numerator) such that $t$ corresponds to the number of nodes in the network.
As can be easily seen from Eq. (\ref{eqn:rate}), our growth algorithm provides two well-known limiting cases.  For $p=1$ we retrieve the preferential attachment growth process \cite{barabasi_albert}. For $p=0$, the network will not add nodes and must asymptotically become a clique of size $N_0$. In between, for $p<< 1$, the second term dominates, which renders the network more dense, and produces the large deviation from power-law structure in the distribution tail.

\begin{figure}[h!]
  \centering
    \includegraphics[height=50mm]{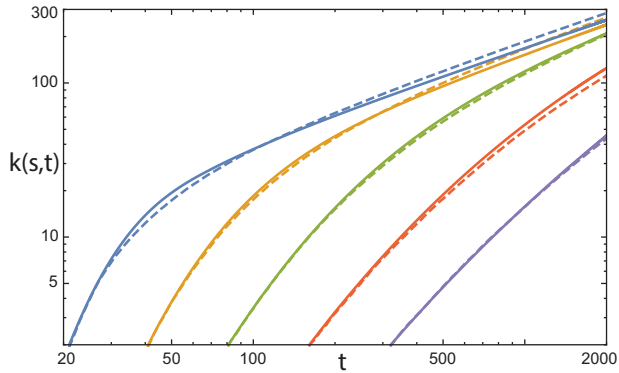}
    \caption{Comparison: Primary model / semi-analytical description. Degree evolution $k(s,t)$ of nodes entering the network at $s=21,41,81,161,321$. Mean of $10^3$ primary model realizations (dashed), compared with numerical integration of Eq. (\ref{eqn:rate}) (solid).}
    \label{fig:trajectories}
\end{figure}

\begin{figure}[h!!!!!!!!]
  	\includegraphics[width=80mm]{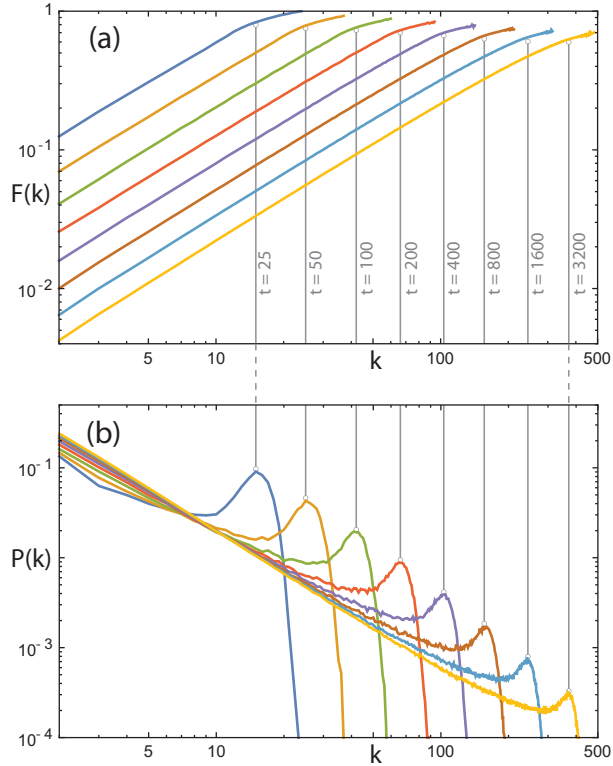}
    \caption{Relation between power-law deviation hump and saturation function: a) Mean field  saturation $F(k,t)$, b) mean of the degree distribution. Data set: $10^3$ network realizations for given time $t$ using $p=\frac{1}{24}$. Vertical grey lines are visual aids. The figure indicates the disappearance of the hump structure in the thermodynamic limit.}
        \label{fig:saturation}

\end{figure}

To demonstrate the validity of our mean-field approximation, we compare the node degree evolution obtained from a $4^{th}$ order Runge-Kutta integration of Eq. (\ref{eqn:rate}) using our approximation for $F(k,t)$ (see below), against the averaged result from $10^3$ realizations of the primary model. As the result, an approximate power law scaling clearly emerges at early evolution stage, and an upper bound to the envelope of node degrees emerges for longer evolution time $t$ necessary to attain larger network sizes (cf. Fig. \ref{fig:trajectories}, where the results of the semi-analytical description are based on exponents and prefactors from an approximation of the results of  Fig. \ref{fig:saturation}a) via Eq. (\ref{2powerlaws})). 
$F(k,t)$ has a very regular behavior in both variables $(k,t)$ (Fig. \ref{fig:saturation}a) and is accompanied by a node degree distribution $P(k)$ as found for our primary model (Fig. \ref{fig:saturation}b). Over a large range, we can approximate $F(k,t)$ by a power law for small $k$, and by a second power law at large $k$:
\begin{equation}
F(k,t) \approx 
\left\{
	\begin{array}{ll}
		t^\alpha k^\beta  & \mbox{if } k \leq k_c \\
		k^\gamma \frac{t^{\alpha}{k_c}^{\beta}}{{k_c}^{\gamma}}& \mbox{if } k > k_c
	\end{array}
\right\} \, ,
\label{2powerlaws}
\end{equation}
where $k_c \sim t^\lambda$, and the fractional term for $k>k_c$ simply makes $F(k,t)$ continuous at $k_c$. The exponents $\alpha,\beta,\gamma,\lambda$ will vary according to the choice of algorithm parameter $p$, where 
$0 < \lambda < 1$: i.e. $1 < k_c <t$. 
In accordance with Fig. \ref{fig:saturation}a), the following observations can be made: First, $\gamma < \beta$ (the exponent of the power law fit decreases as $k$ crosses $k_c$). Second, $F(t-1,t)=1$, since $t-1$ is the maximum possible node degree at time $t$ (achieved in Fig. \ref{fig:saturation}a) for $t=25$ only). 
Similarly, as $p \rightarrow 0$, $F(k,t) \rightarrow 1$, (the network will tend toward a clique, where all possible connections already exist).  When $p = 1$, $F(k,t)$ ceases to be relevant. Finally,
for any $p\in (0,1)$, as $t \rightarrow \infty$, $F(k,t) \rightarrow 0$, since the number of inside edges added at each time-step approximates a constant value, so the network becomes increasingly sparse.

We can use $F(k,t)$ to infer the generated unnormalized degree probability distribution, $N(k,t)$ as follows. Starting from the continuity equation
, we may write 
\begin{equation}
\frac{\partial}{\partial t}N(k,t) =-\frac{\partial}{\partial k}\bigl( N(k,t) \frac{\partial k}{\partial t} \bigr) + \delta_{m,k} \,,
\label{eqn:master}
\end{equation}
where $\frac{\partial k}{\partial t}$ is given by Eq. (\ref{eqn:rate}), and the Kronecker delta function has been included to account for the addition of outside nodes.
By differentiating Eq. (\ref{eqn:rate}), we notice that Eq. (\ref{eqn:master}) contains the product of $k$ and the derivative of the saturation function $F$:
\begin{equation}
\frac{\partial }{\partial k}\frac{\partial k}{\partial t} = a_0 + a_1 - a_1\bigl(k \frac{\partial}{\partial k} F(k,t) +F(k,t)\bigr)\,,
\label{eqn:derivative}
\end{equation}
where $a_0:=\frac{m}{2 K(t)}$, $a_1:=\frac{(1-p)}{p K(t)}$.  The form of $F(k,t)$ implies that a sharp change should occur in the solutions of Eq. (\ref{eqn:derivative}) around $k_c$.  Indeed, a comparison between $P(k,t)$ and $F(k,t)$ (Fig. \ref{fig:saturation}) supports this suggestion. Thus, we hold the properties of the saturation function $F(k,t)$  responsible for the form of the deviation of $P(k,t)$ from the ideal power law. 

\section*{Discussion}

Examples of edge saturation network growth emerge from the fundamental situation where the state of a physical system is described by a symbol, and where time acting on the states leads to a description in terms of a language (symbolic dynamics and formal languages \cite{grassberger,cvitanovic,hao,grebogi,stoop1,stoop2,stoop3,klages}, natural languages). Starting with a finite number of $N_0$ states, observations of the system in time yield sequences of states, that define links on a graph between nodes (states), which implies that more important or more versatile nodes will have more links. As such a network evolves for a finer description, two processes may occur: 1) adjacencies are established between previously unconnected nodes (preferentially between more versatile ones); 2) a new node is added and connected preferentially to already highly connected nodes. Evidently, in many networks there will, however, be a limitation on the number of edges that can be hosted by a given node.

The \emph{Drosophila} courtship body language of 37 fundamental  behavioral states \cite{stoop4,stoop5} and its network is an example of such a process. The states are fundamental in the sense that each act could, from the view of the physics of body motion, be followed by any other act. Some transitions, however, are generally not taken, leading to edges missing. Well-defined connected sub-networks characterize a chosen courtship partner's class, according to which protagonists can be distinguished (male, female  (virgin, mature, mated), fruitless). Within these bounds, courtship exploits the available expression space, corroborating the view that it might advertise individual properties of the sender, into the eyes of a courtship partner \cite{stoop5,stoop6}.
To compare our network growth algorithm with the data from male-female interaction, we grow the network until the number of nodes (symbols) is depleted, with $p$ chosen so that on average the number of edges matches that of the courtship network. A comparison -without further fitting- exhibits that the two degree distributions match extremely well and that the proposed generating algorithm is very specific (Fig.  \ref{fig:varym2}).

Our paradigm may also appear in the guise of an equilibrium condition in the following sense. Complex networks in physics or in biology are often constrained to maintain some 'average' conditions. As soon as (possibly: self-enhancing) node interaction sets in, this needs to be balanced by homeostasis, i.e. a competitive, counter-balancing mechanism that weakens other connections of the same node to the network \cite{Boccaletti2}. In the neural networks domain, a closely related principle is known as `Hebbian learning' \cite{Hebb}. Self-organized Hebbian-learning \cite{landis} in the super-paramagnetic \cite{jcics} phase of ensembles has been proven a reliable and efficient way of clustering that does away with convexity requirements of cluster borders \cite{bioinformatics}.
A very similar approach has also been used as a synchronization model for coupled oscillators, where the oscillators' struggle to synchronize is expressed by competing connection strengths  $w_{ij}$ that evolve according to the dynamical update rule $\frac{dw_{ij}}{dt} = s_{ij} - w_{ij}\bigl(\sum_{(i,k) \in E}s_{ik}\bigr)$  \cite{Boccaletti2}, where $s_{ij}$ measures the pairwise oscillator synchrony. The resulting distribution of $w_{ij}$ has been shown to tend for intermediate coupling strengths towards a hump-terminated power-law (cf. Fig. \ref{fig:varyp0}a). This dynamical law expresses the limited resources available for the local wiring around each node, which in our model is encoded in the probability $p$ ruling the edge saturation.
We envisage that also avalanche distributions of the typical form of Fig. 2a) could be understood similarly \cite{hans1}. 

Many interesting real-world phenomena dwell on the mesoscale. In social networks, the largest scale is relevant, e.g., for the study of disease and rumor spreading, but more subtle social dynamics happens within the community structures \cite{girvan_newman,palla}. Our results suggest that a large class of systems can be formulated as growing along simple principles, similar and in addition to preferential attachment. The sets of $m$, $p$ parameters needed to recover an experimental distribution, i.e. the violation of the ideal power law on the macroscopic scale, provides us with an insight about the local mesoscale structures present in the network. In this way, starting from non-ideal power law distributions of complex networks, an avenue opens towards the identification and understanding of interesting mesoscale real-world phenomena in physics.

Work supported by the Swiss National Science Foundation (Grant 200021-153542/1 to R.S.).

\section*{Author contributions}
R.S. and T.L. designed the research, T.L. and F.G. carried out the analysis, R.S. wrote the manuscript.
\section*{Additional information}
Competing financial interests: The authors declare no competing financial interests.


\begin{thebibliography}{}

\bibitem{Boccaletti_review} S. Boccaletti, V. Latora, Y. Moreno, M. Chavez, and D.-U. Hwang, Complex networks: structure and dynamics,
Phys. Rep. \textbf{424}, 175 (2006).

\bibitem{barabasi_review} R. Albert and A.-L. Barab\'{a}si, Statistical mechanics of complex networks, Rev. Mod. Phys. {\bf 74}, 47 (2002).

\bibitem{shlomo_book}R. Cohen and S. Havlin, \textit{Complex networks: structure, robustness and function} (Cambridge University Press, 2010).

\bibitem{barabasi_albert}A.-L. Barab\'{a}si and R. Albert, Emergence of scaling in random networks, Science {\bf 286}, 509 (1999).

\bibitem{stanley1}L.A.N. Amaral, A. Scala, M. Barth\'el\'emy, and H.E. Stanley, Classes of small-world networks, Proc. Natl. Acad. Sci. U.S.A. \textbf{97}, 11149 (2000).

\bibitem{stanley2}S. Mossa, M. Barth\'el\'emy, H.E. Stanley, and L.A.N. Amaral, Truncation of power law behavior in ``scale-free'' network models due to information filtering, Phys. Rev. Lett. \textbf{88}, 138701 (2002).

\bibitem{dorogovtsev_mendes_wordwebs} S.N. Dorogovtsev and J.F.F. Mendes, Language as an evolving word web, Proc. R. Soc. Lond. B {\bf 268}, 2603 (2001).

\bibitem{Boccaletti2} S. Assenza, R. Guti\'{e}rrez,  J. G\'{o}mez-Garde\~{n}es, V. Latora, and S. Boccaletti, Emergence of structural patterns out of synchronization in networks with competitive interactions, Sci. Rep. \textbf{1}, 99 (2011).


\bibitem{herman1} C.W. Eurich, J.M. Herrmann, and U.A. Ernst, Finite-size effects of avalanche dynamics, Phys. Rev. E {\bf 66}, 066137 (2002).  

\bibitem{herman2}A. Levina, J.M. Herrmann, and T. Geisel, Dynamical synapses causing self-organized criticality in neural networks, Nat. Phys. {\bf 3}, 857 (2007). 

\bibitem{hans1} L. de Arcangelis, F. Lombardi, and H.J. Herrmann, Criticality in the brain, J. Stat. Mech. {\bf 3}, P03026 (2014). 







\bibitem{Camacho} I. Yanai, C.J. Camacho and C. DeLisi, Predictions of gene family distributions in microbial genomes: evolution by gene duplication and modification, Phys. Rev. Lett. {\bf 85}, 2641 (2000).

\bibitem{GrossPRL} C. I. Del Genio, T. Gross, and K.E. Bassler, All scale-free networks are sparse, Phys. Rev. Lett. {\bf 107}, 178701 (2011).

\bibitem{dorogovtsevPRE} S.N. Dorogovtsev, J.F.F. Mendes, and A.N. Samukhin, Size-dependent degree distribution of a scale-free growing network, Phys. Rev. E {\bf 63}, 062101 (2001).

\bibitem{guimaraes} P.R. Guimaraes, M.A.M. de Aguiar, J. Bascompte, P. Jordano, and S.F. dos Reis, Random initial condition in small Barabasi-Albert networks and deviations from the scale-free behavior, Phys. Rev. E {\bf 71}, 037101 (2005).

\bibitem{waclaw}B. Waclaw and I.M. Sokolov, Finite-size effects in Barab\'asi-Albert growing networks, Phys. Rev. E {\bf 75}, 056114 (2007).



\bibitem{stoop4} R. Stoop and B. Arthur, Periodic orbit analysis demonstrates genetic constraints, variability, and switching in Drosophila courtship behavior, Chaos {\bf 18}, 023123 (2008).

\bibitem{stoop5} R. Stoop and J. Joller, Mesocopic Comparison of Complex Networks Based on Periodic Orbits, Chaos {\bf 21}, 016112 (2011).


\bibitem{dorogovtsev_mendes_book} S.N. Dorogovtsev and J.F.F. Mendes, {\it Evolution of Networks} (Oxford University Press, Oxford, 2003).

\bibitem{grassberger} P. Grassberger and H. Kantz, Generating partitions for the dissipative H\'enon map, Phys. Lett. A \textbf{113}, 235 (1985).

\bibitem{cvitanovic}P. Cvitanovi\'c, G.H. Gunaratne, and I. Procaccia, Topological and metric properties of H\'enon-type strange attractors, Phys. Rev. A \textbf{38}, 1503 (1988).

\bibitem{hao} H. Bai-Lin, {\it Elementary Symbolic Dynamics and Chaos in Dissipative Systems} (World Scientific, Singapore, 1989).

\bibitem{stoop3} R. Stoop, Bivariate thermodynamic formalism and anomalous diffusion, Phys. Rev. E {\bf 49}, 4913 (1994).

\bibitem{stoop1} R. Stoop and J. Parisi, Evaluation of probabilistic and dynamical invariants from finite symbolic substrings-comparison between two approaches, Physica D {\bf 58}, 325 (1992).

\bibitem{stoop2}R. Stoop, Phase transitions in the approximated and asymptotic generalized entropy spectrum of a nonhyperbolic system, Phys. Rev. A {\bf 46}, 7450 (1992).

\bibitem{grebogi}Y.-C. Lai, E. Bollt, and C. Grebogi, Communicating with chaos using two-dimensional symbolic dynamics, Phys. Lett. A {\bf 255}, 75 (1999).

\bibitem{klages} R. Klages, {\it Microscopic chaos, fractals and transport in non-equilibrium statistical mechanics} (World Scientific, Singapore, 2007).

\bibitem{stoop6} R. Stoop, P. N\"{u}esch, R. L. Stoop, and L.A. Bunimovich, At grammatical faculty of language, flies outsmart men, PLoS ONE {\bf 8}, e70284 (2013).

\bibitem{Hebb} D. Hebb, {\it The Organization of Behavior} (Wiley \& Sons, New York, 1949).

\bibitem{landis} F. Landis, T. Ott, and R. Stoop, Hebbian self-organizing integrate-and-fire networks for data clustering, Neur. Comp. {\bf 22}, 273 (2010).

\bibitem{jcics} T. Ott, A. Kern, A. Schuffenhauer, M. Popov, P. Acklin, E. Jacoby, and R. Stoop, Sequential superparamagnetic clustering for unbiased classification of high-dimensional chemical data, J. Chem. Inf. Comput. Sci. {\bf 44}, 1358 (2004).
\bibitem{bioinformatics} F. Gomez, R.L. Stoop, and R. Stoop, Universal dynamical properties preclude standard clustering in a large class of biochemical data,
Bioinformatics {\bf 30}, 2486 (2014).

\bibitem{girvan_newman} M. Girvan and M.E.J. Newman, Community structure in social and biological networks, Proc. Natl. Acad. Sci. U.S.A. {\bf 99}, 7821 (2002).

\bibitem{palla} G. Palla, I. Der\'{e}nyi, I. Farkas, and T. Vicsek, Uncovering the overlapping community structure of complex networks in nature and society, Nature {\bf 435}, 814 (2005).

\end{thebibliography}
\end{document}